\PassOptionsToPackage{unicode}{hyperref}
\PassOptionsToPackage{hyphens}{url}
\PassOptionsToPackage{dvipsnames,svgnames,x11names}{xcolor}
\documentclass[
  12pt]{article}
\usepackage{comment}
\usepackage{amsmath,amssymb}
\usepackage{iftex}
\usepackage{bm}
\usepackage{enumitem}
\usepackage{multirow}
\usepackage{lscape}
\usepackage{makecell}
\usepackage{diagbox}
\usepackage{tikz}
\usepackage{xcolor}
\usepackage{ulem}
\definecolor{myPink}{RGB}{251,97,215}
\DeclareRobustCommand{\ggcircle}{%
	\tikz[baseline=-0.6ex]{
		\draw[myPink!70, line width=0.5pt] (0,0) circle (0.30ex);
		\draw[myPink!70, line width=0.7pt] (-1.2ex,0) -- (1.2ex,0);
	}%
}
\ifPDFTeX
  \usepackage[T1]{fontenc}
  \usepackage[utf8]{inputenc}
  \usepackage{textcomp} 
\else 
\fi
\usepackage{lmodern}
\ifPDFTeX\else  
\fi
\IfFileExists{upquote.sty}{\usepackage{upquote}}{}
\IfFileExists{microtype.sty}{
  \usepackage[]{microtype}
  \UseMicrotypeSet[protrusion]{basicmath} 
}{}
\makeatletter
\@ifundefined{KOMAClassName}{
  \IfFileExists{parskip.sty}{%
    \usepackage{parskip}
  }{
    \setlength{\parindent}{0pt}
    \setlength{\parskip}{6pt plus 2pt minus 1pt}}
}{
  \KOMAoptions{parskip=half}}
\makeatother
\usepackage{xcolor}
\makeatletter
\ifx\paragraph\undefined\else
  \let\oldparagraph\paragraph
  \renewcommand{\paragraph}{
    \@ifstar
      \xxxParagraphStar
      \xxxParagraphNoStar
  }
  \newcommand{\xxxParagraphStar}[1]{\oldparagraph*{#1}\mbox{}}
  \newcommand{\xxxParagraphNoStar}[1]{\oldparagraph{#1}\mbox{}}
\fi
\ifx\subparagraph\undefined\else
  \let\oldsubparagraph\subparagraph
  \renewcommand{\subparagraph}{
    \@ifstar
      \xxxSubParagraphStar
      \xxxSubParagraphNoStar
  }
  \newcommand{\xxxSubParagraphStar}[1]{\oldsubparagraph*{#1}\mbox{}}
  \newcommand{\xxxSubParagraphNoStar}[1]{\oldsubparagraph{#1}\mbox{}}
\fi
\makeatother

\usepackage{longtable,booktabs,array}
\usepackage{calc} 
\usepackage{etoolbox}
\makeatletter
\patchcmd\longtable{\par}{\if@noskipsec\mbox{}\fi\par}{}{}
\makeatother
\IfFileExists{footnotehyper.sty}{\usepackage{footnotehyper}}{\usepackage{footnote}}
\makesavenoteenv{longtable}
\usepackage{graphicx}
\makeatletter
\def\maxwidth{\ifdim\Gin@nat@width>\linewidth\linewidth\else\Gin@nat@width\fi}
\def\maxheight{\ifdim\Gin@nat@height>\textheight\textheight\else\Gin@nat@height\fi}
\makeatother
\setkeys{Gin}{width=\maxwidth,height=\maxheight,keepaspectratio}
\makeatletter
\def\fps@figure{htbp}
\makeatother

\makeatletter
\@ifpackageloaded{caption}{}{\usepackage{caption}}
\AtBeginDocument{%
\ifdefined\contentsname
  \renewcommand*\contentsname{Table of contents}
\else
  \newcommand\contentsname{Table of contents}
\fi
\ifdefined\listfigurename
  \renewcommand*\listfigurename{List of Figures}
\else
  \newcommand\listfigurename{List of Figures}
\fi
\ifdefined\listtablename
  \renewcommand*\listtablename{List of Tables}
\else
  \newcommand\listtablename{List of Tables}
\fi
\ifdefined\figurename
  \renewcommand*\figurename{Figure}
\else
  \newcommand\figurename{Figure}
\fi
\ifdefined\tablename
  \renewcommand*\tablename{Table}
\else
  \newcommand\tablename{Table}
\fi
}
\@ifpackageloaded{float}{}{\usepackage{float}}
\floatstyle{ruled}
\@ifundefined{c@chapter}{\newfloat{codelisting}{h}{lop}}{\newfloat{codelisting}{h}{lop}[chapter]}
\floatname{codelisting}{Listing}

\makeatother
\makeatletter
\@ifpackageloaded{caption}{}{\usepackage{caption}}
\@ifpackageloaded{subcaption}{}{\usepackage{subcaption}}
\makeatother

\ifLuaTeX
  \usepackage{selnolig}  
\fi
\usepackage[]{natbib}
\usepackage{bookmark}

\IfFileExists{xurl.sty}{\usepackage{xurl}}{} 
\hypersetup{
  pdftitle={Title},
  pdfauthor={Author 1; Author 2},
  pdfkeywords={3 to 6 keywords, that do not appear in the title},
  colorlinks=true,
  linkcolor={blue},
  filecolor={Maroon},
  citecolor={Blue},
  urlcolor={Blue},
  pdfcreator={LaTeX via pandoc}}

\def\bX{\mathbf{X}}
\def\bx{\mathbf{x}}

\def\bH{\mathbf{H}}
\def\bM{\mathbf{M}}

\def\bD{\mathbf{D}}

\def\bz{\mathbf{z}}

\def\bbeta{\boldsymbol{\beta}}

\def\bbeta{\boldsymbol{\beta}}

\newtheorem{theorem}{Theorem}

\newtheorem{remark}{Remark}
\newtheorem{proposition}{Proposition}
\newtheorem{assumption}{Assumption}

\newcommand{\anon}{1}

\begin{document}

\def\spacingset#1{\renewcommand{\baselinestretch}%
{#1}\small\normalsize} \spacingset{1}


\if1\anon
{
  \title{\bf Causal Inference for Heterogeneous Extreme Quantiles with Heavy-Tailed Outcomes}
  \author{Xiaorui Wang\hspace{.2cm}\\
    School of Mathematics and Statistics,\\ Nanjing University of Information Science and Technology\\
    and \\
    Juan-Juan Cai \\
    Department of Econometrics and Data Science,\\ Vrije Universiteit Amsterdam\\
and \\
Huixia Judy Wang \\
Department of Statistics, Rice University\\
and \\
Jian Qing Shi \\
Guangdong Provincial/Zhuhai Key Laboratory of\\ Interdisciplinary Research and Application for Data Science,\\ Beijing Normal-Hong Kong Baptist University\\
and \\
Yanlin Tang\thanks{
	The authors gratefully acknowledge financial support from the National Natural Science Foundation of China (No. 12371265 and 12271239), Fundamental and Interdisciplinary Disciplines Breakthrough Plan of the Ministry of Education of China (No. JYB2025XDXM904), Natural Science Foundation of Shanghai (No. 24ZR1455200), and National Science Foundation (No. DMS-2436216 and DMS-2426174).} \\
Key Laboratory of Advanced Theory and\\ Application in Statistics and Data Science - MOE,\\ School of Statistics, East China Normal University}
\date{}
  \maketitle
} \fi

\if0\anon
{
  \bigskip
  \bigskip
  \bigskip
  \begin{center}
    {\LARGE\bf Causal Inference for Heterogeneous Extreme Quantiles with Heavy-Tailed Outcomes}
\end{center}
  \medskip
} \fi

\bigskip
\begin{abstract}

We propose a framework for estimating conditional extreme quantile treatment effects (CEQTEs) in observational studies with heavy-tailed outcomes. Our procedure first estimates intermediate conditional quantiles using inverse-probability-weighted (IPW) quantile regression and then extrapolates them to extreme levels using extreme value theory. Under a linear conditional quantile model, we show that the conditional and marginal distributions of each potential outcome share a common extreme value index (EVI), motivating two complementary Hill-type EVI estimators based on conditional and marginal information, respectively.
On the theoretical front, we introduce an IPW tail quantile score process that bridges regression quantile score processes and uniform tail empirical processes while accounting for treatment assignment. We establish its functional weak convergence under mild regularity conditions, without requiring a max-domain-of-attraction condition. This result provides the probabilistic foundation for the asymptotic analysis of the proposed CEQTE estimators. Simulation studies demonstrate favorable finite-sample performance, and an application to NLSY79 data reveals substantial heterogeneity in the effect of college education on extremely high hourly wages across confounder-defined subpopulations.
\end{abstract}

\noindent%
{\it Keywords:} Conditional extreme quantile treatment effect; Extrapolation; Extreme value index; IPW tail quantile score process
\vfill

\newpage
\spacingset{1.8} 

\section{Introduction}\label{sec-intro}

Personalized treatment and intervention have attracted widespread attention in many scientific and policy applications. Heterogeneous treatment effects (HTEs) provide a powerful framework for understanding how the causal effect of a treatment varies conditional on  confounders. While most existing studies focus on average or regular quantile treatment effects, causal effects at extreme quantile levels are often of particular interest, especially when outcomes exhibit heavy-tailed behavior. For example, in the education returns dataset considered in Section \ref{Real data}, analyzing the causal effect of college education on extremely high hourly wages across individuals with different characteristics can provide valuable insights into income inequality and targeted policy interventions. The HTEs at extreme quantiles characterize how a treatment changes the upper tail of the outcome distribution while simultaneously capturing the heterogeneity of these effects across confounder subpopulations.

Widely studied measures of HTEs include the conditional average treatment effect (CATE) and the conditional quantile treatment effect (CQTE). 
The CATE characterizes treatment effect heterogeneity in terms of the conditional mean, while the CQTE extends this framework by evaluating how treatment effects vary across different parts of the conditional outcome distribution. A large body of literature has investigated the estimation and inference of CATE under various settings. \cite{Abrevaya:Hsu:Lieli} discussed the nonparametric identifiability and the inverse probability weighted estimation of  CATE under the unconfoundedness assumption. \cite{Jacob:2021} developed CATE estimators based on machine learning methods, and \cite{Huang:Yang:2020} provided a robust inference of the CATE using dimension reduction. Recent developments in statistical methods for assessing treatment effect heterogeneity in randomized clinical trials and observational studies are summarized in \cite{lipkovich2024modern}. Compared with CATE, CQTE provides a more comprehensive characterization of treatment effect heterogeneity by allowing treatment effects to vary across the conditional outcome distribution. 
For example, in modern drug development, it is important to identify sub-populations in which a treatment is most beneficial or harmful \citep{Ma:Huang:2017}; in the empirical study of infant birth weights, \cite{Cai2021} demonstrated that the effect of maternal smoking varies across quantile levels conditional on predetermined confounders. Recent methodological advances have established theoretical guarantees for CQTE estimation under increasingly complex settings. \cite{zhou2022role} studied the asymptotic normality of low-dimensional CQTE estimators based on propensity scores,   \cite{Giessing:Wang:2023} developed inference procedures for CQTE with high-dimensional confounders, and \cite{qiu2025heterogeneous} proposed an estimation approach based on convolution-smoothed quantile regression and orthogonal random forests. Nevertheless, existing HTE methods mainly focus on  treatment effects at the conditional mean or regular quantile levels, leaving treatment effect heterogeneity in the extreme tails of the outcome distribution largely unexplored. This gap motivates the development of extreme quantile inference methods based on EVT.

Recent studies have begun to investigate treatment effects at extreme quantile levels, motivated by applications involving rare but consequential outcomes. \citet{Zhang2018} provides one of the early studies on this problem. Using inverse-probability-weighted quantile methods, \citet{Zhang2018} develops estimation and inference for marginal extremal quantile treatment effects under different types of tail behavior. Building on this work, \citet{Deuber:Li:Engelke:Maathuis:2021} focus on the heavy-tailed setting and exploit extreme value theory (EVT) to extrapolate from intermediate to more extreme quantile levels. Relatedly, \citet{huang2024estimation} study marginal extremal quantile treatment effects with a continuous treatment, extending the literature beyond the binary-treatment setting.  \citet{chen2025estimation} consider expected-shortfall-type treatment effects and employ the Fissler--Ziegel loss to estimate parameters in linear quantile models.  

In this paper, we propose a framework for estimating the conditional extreme quantile treatment effect (CEQTE) for heavy-tailed outcomes. The CEQTE characterizes treatment heterogeneity in the extreme tails conditional on confounders. We first estimate intermediate conditional quantiles for the treatment and control groups 
and then extrapolate these estimates to extreme levels through EVT. A central methodological challenge is the stable estimation of the extreme value index (EVI), which is crucial for EVT-based extrapolation. In our causal setting, this task is particularly difficult because tail observations are inherently scarce and, for each individual, one of the two potential outcomes is unobserved. We address this by showing that, under a linear conditional quantile model, the conditional and marginal distributions of each potential outcome share a common EVI. This result motivates two complementary Hill-type estimators: a conditional estimator based on estimated conditional tail quantiles and a marginal estimator that pools information across covariate values. Our simulations reveal that the marginal estimator is more stable in multivariate covariate settings and with limited sample sizes, a setting commonly encountered in applications, whereas the conditional estimator  becomes increasingly competitive and eventually outperforms the marginal estimator as the sample size grows. 
As a key theoretical tool, we introduce an
\textit{inverse-probability-weighted (IPW) tail quantile score process}. 
This process plays a dual role, combining features of the regression rank-score process in quantile regression with those of the uniform tail empirical process in extreme value theory, while additionally incorporating inverse-probability weighting for causal inference. We establish the functional weak convergence of this process (see Theorem~\ref{thm: AN_quantilescore}) under fairly general conditions that do not require the conditional outcome distribution to belong to a max-domain of attraction.

Our framework therefore bridges two previously separate strands of the
literature: causal inference for heterogeneous treatment effects and extreme
conditional quantile estimation. Unlike existing work on marginal extremal
quantile treatment effects, the CEQTE captures treatment-effect heterogeneity
across covariate profiles; unlike conventional extreme conditional quantile
regression, our target is a causal contrast between potential-outcome
distributions.

We illustrate the practical relevance of the proposed framework using data from
the National Longitudinal Survey of Youth 1979 (NLSY79) to study the effect of
college education on extremely high hourly wages. The estimated CEQTE exhibits
substantial heterogeneity across covariate profiles: the effect of college
education is considerably larger for individuals residing in the Northeast and
increases markedly with cognitive ability, as measured by ASVAB scores, while
varying relatively little with family income. The estimated treatment effect
also becomes larger at more extreme quantile levels, which is consistent with the heavier upper tail of wages among college-educated individuals. Importantly,
these conditional effects differ substantially from the corresponding marginal
extreme quantile treatment effects, illustrating how marginal analyses can mask
economically meaningful heterogeneity in the extreme tail.

The rest of the article is organized as follows.
Section \ref{Method} introduces the proposed extrapolation estimation of the CEQTE and two Hill-type estimators for the EVI.
Section \ref{Theory} establishes the asymptotic properties of the IPW tail quantile score process and the proposed EVI and CEQTE estimators.
Sections \ref{Simulation} and  \ref{Real data} evaluate the proposed method through  simulation studies and an empirical analysis of the causal effect of college education on the upper tail of hourly wages, respectively.
Section \ref{Discussion} concludes with discussions and future directions.
Technical proofs and additional implementation details are provided in the online Supplementary Materials.

\section{Model and Estimation}\label{Method}

Let $Y$ be the observed outcome, $\bX = (1,X_{1}, \ldots, X_{d-1})^{\top}$ the $d$-dimensional confounder vector, 
and $T\in\{0,1\}$ the treatment indicator, where $T = 1$ denotes treatment and 0 for control. Let $Y(1)$ and $Y(0)$ denote the potential outcomes under treatment and control, respectively.
Since only one of the two potential outcomes is observed for each individual, identification relies on the following standard conditions.

\begin{assumption}[Conditions for  potential outcomes framework]
	\label{asmp: Assumption 1}
	\leavevmode
	\begin{enumerate}[label=(\roman*)]
		\item Consistency: $Y = TY(1)+(1-T)Y(0)$.
		\item Strongly ignorable treatment assignment (unconfoundedness): $\{Y(1),Y(0)\}\perp T\mid\bX$ where $\perp$ denotes independence.
		\item Positivity: $0<c_{1}\leq\pi(\bX)\leq c_{2}<1$ for some $c_{1},c_{2}\in (0,1)$, where $\pi(\bX)=P(T=1|\bX)$ is the propensity score.
	\end{enumerate}
\end{assumption}

For $j=0,1$, let $F_{Y(j)|{\bx}}(\cdot)$ denote the conditional cumulative distribution function of $Y(j)$ given ${\bX}=\bx$, and $Q_{Y(j)|\bx}(\cdot)$ be the corresponding conditional quantile function. Let $\mathcal{D}=\{(Y_{s},\bX_{s},T_{s}),s=1,\cdots,n\}$ be a random sample of $(Y,\bX,T)$.
Our main objective is to estimate the conditional extreme quantile treatment effect at level $p_n$ (denoted $p_n$-CEQTE),
$$
\Delta(p_n|{\bx_0}) = Q_{Y(1)|{\bx_0}}(p_n)-Q_{Y(0)|{\bx_0}}(p_n),
$$
where $\bx_0$ denotes the specific confounder value at which the treatment effect is of interest.
We consider the extreme quantile regime where $p_n\rightarrow 1$ and $n(1-p_n)\rightarrow c \in[0, \infty)$ as $n \rightarrow \infty$.
The CEQTE captures not only the treatment effect at the upper tail of the outcome distribution, but also reveals heterogeneity across subpopulations defined by confounders $\bx_0$.

To estimate $\Delta(p_n\mid\bx_0)$, we need estimate the conditional potential-outcome quantiles $Q_{Y(j)\mid\bx_0}(p_n)$ for $j=0,1$. Our estimation strategy relies on two structural assumptions on the upper-tail conditional quantiles. We first impose a linear quantile structure, which enables us to estimate the potential-outcome conditional quantiles at intermediate levels. We then introduce a heavy-tail condition, which provides the additional structure needed to extrapolate these intermediate conditional quantiles to the extreme level $p_n$.

\subsection{Intermediate quantile estimation under the linear structure}\label{sec: linear model}

We formalize the linear upper-tail quantile structure in the following assumption.
\begin{assumption}
	\label{asmp: linear}
	Suppose that $\bX$ has compact support $\bD_{\bX}\subset \mathbb{R}^{d}$. For simplicity, we take
	$
	\bD_{\bX}=\{1\}\times[0,1]^{d-1},
	$
	where the first component of $\bX$ corresponds to an intercept. Let $0<\tau_c<1$ be a fixed constant, possibly close to one. For every $\tau\in[\tau_c,1)$ and $j\in\{0,1\}$, there exists
	\[
	{\bm\beta}_{j}(\tau)
	=
	\bigl(
	\beta_{j,0}(\tau),
	\beta_{j,1}(\tau),
	\ldots,
	\beta_{j,d-1}(\tau)
	\bigr)^{\top}
	\]
	such that, for every $\bx\in\bD_{\bX}$,
	\begin{equation}\label{model1}
		Q_{Y(j)\mid \bx}(\tau)
		=
		\bx^{\top}{\bm\beta}_{j}(\tau),
		\qquad j=0,1.
	\end{equation}
	For $\tau\in(0,\tau_c)$, no restriction is imposed on the conditional quantiles of the potential outcomes.
\end{assumption}

\begin{remark}
Assumption~\ref{asmp: linear} imposes linearity in the confounders while leaving the coefficient vector $\bm\beta_j(\tau)$ unrestricted as a function of $\tau$ over the upper tail. Consequently, \eqref{model1} is not a finite-dimensional parametric specification. For example, the components in the coefficient vector may be constant, bounded, or regularly varying as $\tau\rightarrow1$, allowing the model to accommodate a broad range of upper-tail behaviors; see Section~\ref{Simulation} for examples. At the same time, the linear structure makes it possible to characterize the tail behavior of the conditional and unconditional outcome distributions in a tractable manner; see Proposition~\ref{unconditional}. More flexible models for confounder-dependent extreme quantiles have also been studied, including semiparametric approaches \citep{DS90,WangTsai2009,CDD05,you2019} and nonparametric approaches \citep{Daouia2013,Velthoenetal2019,GardesStupfler2019}. 
\end{remark}

Based on Assumption~\ref{asmp: linear}, we apply linear quantile regression to estimate the quantile coefficient vector
${\bm\beta}_{j}(\cdot)$ at intermediate quantile levels. Let $k_n$ and $m_n$ be two intermediate sequences such that $k_n\rightarrow \infty$ and $k_n/n\rightarrow 0$ as $n\rightarrow \infty$, with $m_n < k_n$. We aim to estimate ${\bm\beta}_{j}(\tau_n)$ for 
\[
\tau_n \in \left\{1-\frac{k_n}{n},\, 1-\frac{k_n-1}{n},\, \ldots,\, 1-\frac{m_n}{n}\right\}.
\]
Each $\tau_n$ corresponds to a quantile level in the tail region, but one at which a sufficient number of observations remain above the threshold.

In observational studies, however, the subsample $\{(\bX_s, Y_s) : T_s = j\}$ cannot be directly used to estimate ${\bm\beta}_{j}(\tau_n)$ because the distribution of confounders generally differs across treatment and control groups, leading to an imbalanced design. A standard remedy is inverse probability weighting (IPW). As shown in \citet[Lemma G.2]{Deuber:Li:Engelke:Maathuis:2021}, for any measurable function $g$,
\[
E[g\{Y(1)\}]=E\left\{ g(Y)\frac{T}{\pi(\bX)}\right\}.
\]
Applying this identity to the quantile regression problem yields the estimator
\begin{eqnarray}
	\widehat{{\bm\beta}}_{j}\left(\tau_n\right):= \mathop{\arg\min}_{{\bm b}}\sum\limits_{s=1}^{n}\left\{\frac{T_{s}}{\widehat{\pi}(\bX_{s})}\right\}^j\left\{\frac{1-T_{s}}{1-\widehat{\pi}(\bX_{s})}\right\}^{1-j}\rho_{\tau_n}(Y_{s}-{\bX}^{\top}_{s}{\bm b}),   \label{eq: hat_beta}
\end{eqnarray}
where $\rho_{\tau_n}(w)=w\left\{\tau_n-\mathbb{I}(w\leq 0)\right\}$ is the quantile check loss function and $\widehat{\pi}(\bX_{s})$ denotes an estimator of the propensity score $\pi(\bX_s)=P(T_s=1|\bX_s)$.

Among the various methods available for estimating the propensity score, we recommend using the multiply robust estimator of $\pi(\cdot)$, as studied in \cite{Han:2014}; \cite{Han2019}; \cite{Wang:Qin:Song:Tang:2022,Wang:Qin:Tang:Wang:2023} and so on. This estimator offers robustness against misspecification of parametric models while also avoiding the curse of dimensionality typically encountered in nonparametric approaches. We denote the estimator by $\widehat{\pi}(\bX_s)$. 
Its formal definition and implementation details are provided in Section~S2.1 of the online Supplemental Materials.

Accordingly, the conditional intermediate  quantile function can be estimated as
\begin{equation}
	\widehat{Q}_{Y(j)\mid \bx_0}\left(1 - \frac{k_n}{n}\right) := \bx_0^\top \widehat{\bm\beta}_j\left(1 - \frac{k_n}{n}\right), \label{eq: hatQkn}
\end{equation}
where $ \widehat{\bm\beta}_j\left(1 - \frac{k_n}{n}\right)$ is given by \eqref{eq: hat_beta} with $\tau_n=1-k_n/n$.

\subsection{CEQTE estimation under the heavy-tail assumption}\label{sec: heavy tail}

The second structural feature of our model is the heavy-tail assumption on the conditional distributions of the potential outcomes. This additional structure is used for extrapolation from intermediate to extreme quantile levels.
\begin{assumption}
	\label{asmp: heavy-tail}
	For a given $\bx$ in the support of ${\bX}$, there exists a positive $\gamma_{j}(\bx)$ such that 
	for any  $y>0$, 
	\begin{eqnarray}\label{oneorder}
		\lim_{t\rightarrow \infty }\frac{Q_{Y(j)|\bx}\left(1-\frac{1}{vy}\right)}{Q_{Y(j)|\bx}\left(1-\frac{1}{v}\right)}=y^{\gamma_{j}(\bx)}, ~~~ j=0, 1,
	\end{eqnarray}
	that is, $Q_{Y(j)|\bx}\left(1-\frac{1}{v}\right)$ is regularly varying with an index $\gamma_{j}(\bx)$.
	Here,  $\gamma_{j}(\bx)$ is the extreme value index (EVI) of $F_{Y(j)|{\bx}}(\cdot)$. 
\end{assumption}

Assumption~\ref{asmp: heavy-tail} allows the conditional EVI $\gamma_j(\bx)$ to vary with the confounders. Under the linear quantile structure, however, this apparent heterogeneity disappears under a mild condition on the relative tail behavior of the coefficient functions.

\begin{proposition}\label{unconditional}
	Suppose that $\lim_{\tau \rightarrow 1} \frac{\beta_{j,\ell}(\tau)}{\beta_{j,\ell'}(\tau)} = h_{j, \ell\ell'} \in [0, \infty]$ for $0 \leq \ell < \ell' \leq d-1$. Under Assumptions \ref{asmp: linear} and \ref{asmp: heavy-tail}, the marginal distribution of the potential outcome, $F_{Y(j)}$, is heavy-tailed with extreme value index $\gamma_j > 0$, for $j = 0, 1$. Moreover, the EVI of the conditional distribution satisfies $\gamma_j(\bx) \equiv \gamma_j$.
\end{proposition}

Proposition~\ref{unconditional} {reveals a useful structural feature of the linear conditional quantile model that} is important for both interpretation and estimation. Although the levels of the conditional extreme quantiles may vary with $\bx$, their rate of tail decay, as summarized by the EVI, does not. Moreover, the common conditional EVI coincides with that of the marginal distribution of $Y(j)$. This invariance gives rise to two estimation strategies: one based directly on estimated conditional quantiles and another based on the marginal distribution of the potential outcome. 
Throughout the remainder of the paper, we therefore write $\gamma_j$ for the common EVI of both the conditional and marginal distributions of $Y(j)$.
The proof of Proposition~\ref{unconditional} is provided in Section~S1.2 of the online Supplemental Materials.

We next introduce two Hill-type estimators  of $\gamma_j$. The first adapts the classical Hill estimator \citep{Hill:1975} to the conditional quantile framework, whereas the second exploits Proposition~\ref{unconditional} and estimates the same EVI from the marginal distribution of the potential outcome.

Under Assumption~\ref{asmp: heavy-tail}, regular variation yields the
approximation
\[
\frac{1}{k_n}\sum_{i_n=1}^{k_n}
\log\!\left\{
\frac{Q_{Y(j)\mid \bx_0}(1-i_n/n)}
{Q_{Y(j)\mid \bx_0}(1-k_n/n)}
\right\}
\approx
\gamma_j(\bx_0)
\frac{1}{k_n}\sum_{i_n=1}^{k_n}
\log\!\left(\frac{k_n}{i_n}\right)
\approx
\gamma_j(\bx_0).
\]
Replacing the conditional quantiles in this approximation by their estimates
constructed in Section~\ref{sec: linear model} leads to our first estimator
of $\gamma_j$:
\begin{equation} \label{eq: hatgamma1}
	\widehat{\gamma}^{\rm HC}_{j}(\bx_0)
	= \left[k_n-(m_n-1)\left\{1-\log\left( \frac{m_n-1}{k_n}\right)\right\}\right]^{-1}
	\sum_{i_n=m_n}^{k_n}
	\log\!\left\{
	\frac{\bx_0^\top \widehat{\bm\beta}_{j}\!\left(1 - \frac{i_n}{n}\right)}
	{\bx_0^\top \widehat{\bm\beta}_{j}\!\left(1 - \frac{k_n}{n}\right)}
	\right\},
\end{equation}
where $\widehat{\bm\beta}_j(1 - i_n/n)$ is obtained from~\eqref{eq: hat_beta} by setting $\tau_n = 1 - i_n/n, i_n=m_n,\dots,k_n$.

Unlike the classical Hill estimator, here the quantile level cannot be pushed as far as $1 - 1/n$, since the quantile regression coefficients $\bm\beta_j(\tau)$ cannot be reliably estimated at such extreme quantiles (Theorem~\ref{theo: beta}). Consequently, the normalization constant in \eqref{eq: hatgamma1} differs from the usual factor $1/k_n$. A detailed derivation of this scaling factor is provided in Section~S2.2 of the online Supplemental Materials.

Our second estimator exploits the invariance result in Proposition~\ref{unconditional}. Since the EVI of the conditional distribution $F_{Y(j)\mid\bx_0}$ coincides with that of the marginal distribution $F_{Y(j)}$, we may estimate $\gamma_j$ directly from marginal observations. {By pooling tail information across covariate values, this marginal estimator can provide greater finite-sample stability than its conditional counterpart.} We adopt the IPW Hill-type estimator proposed by \citet{Deuber:Li:Engelke:Maathuis:2021},
\begin{eqnarray}\label{eq: hatgamma2}
	\widehat{\gamma}^{{\rm HM}}_{j}=&&\frac{1}{k_{n}}\sum_{\{s: Y_{s}>\widetilde{Q}_{Y(j)}(1-k_n/n)\}}\left\{\frac{T_{s}}{\widehat{\pi}(\bX_{s})}\right\}^j\left\{\frac{1-T_{s}}{1-\widehat{\pi}(\bX_{s})}\right\}^{1-j} \log \frac{Y_s}{\widetilde{Q}_{Y(j)}(1-k_n/n)},
\end{eqnarray}
where the marginal quantile $\widetilde{Q}_{Y(j)}(1-\frac{k_n}{n})$ is estimated non-parametrically through
\begin{eqnarray}
	\widetilde{Q}_{Y(j)}\left(1-\frac{k_n}{n}\right):= \mathop{\arg\min}_{q}\sum\limits_{s=1}^{n}\left\{\frac{T_{s}}{\widehat{\pi}(\bX_{s})}\right\}^j\left\{\frac{1-T_{s}}{1-\widehat{\pi}(\bX_{s})}\right\}^{1-j}\rho_{1-\frac{k_n}{n}}(Y_{s}-q). \label{eq: QYj-uncond}
\end{eqnarray}

We now use these EVI estimators to extrapolate from an intermediate conditional quantile to the target extreme level $p_n$. By regular variation in \eqref{oneorder},
\begin{eqnarray}
	Q_{Y(j)|\bx_0}\left(p_n\right)\approx \Big\{\frac{k_{n}}{n(1-p_n)}\Big\}^{\gamma_{j}}Q_{Y(j)|\bx_0}\left(1-\frac{k_n}{n}\right). \label{eq: QpnApprox}  
\end{eqnarray}
By substituting the estimator $\widehat{Q}_{Y(j)|\bx_0}(1 - \frac{k_n}{n})$ and either of the two estimators of $\gamma_j$ into the approximation in \eqref{eq: QpnApprox}, we obtain an estimator of the conditional quantile at the extreme level $p_n$. Consequently, the conditional extreme quantile treatment effect $\Delta(p_n|\bx_0)$ can be estimated using either method as
\begin{equation}
	\widehat{\Delta}^{\diamond}(p_n|{\bx_0})=\widehat{Q}^{\diamond}_{Y(1)|\bx_0}\left(p_n\right)-\widehat{Q}^{\diamond}_{Y(0)|\bx_0}\left(p_n\right),\quad \diamond=\text{HC}, \text{HM},
	\label{eq: hatDelta}
\end{equation}
with 
\begin{equation*}
	\widehat{Q}^{\diamond}_{Y(j)|\bx_0}\left(p_n\right)=\Big\{\frac{k_{n}}{n(1-p_n)}\Big\}^{\widehat\gamma^{\diamond}_{j}}\widehat Q_{Y(j)|\bx_0}\left(1-\frac{k_n}{n}\right),
\end{equation*}
where $\widehat Q_{Y(j)|\bx_0}\left(1-\frac{k_n}{n}\right)$ is defined in \eqref{eq: hatQkn}; $\widehat\gamma^{\text{HC}}_{j}(\bx_0)$ and $\widehat\gamma^{\text{HM}}_{j}$ are defined in \eqref{eq: hatgamma1} and \eqref{eq: hatgamma2}, respectively.

\section{Large Sample Properties}\label{Theory}

\subsection{Weak convergence of the IPW tail quantile score process} 
\label{sec: asym score process}
The asymptotic analysis developed in this section is built upon the
\textit{inverse-probability-weighted (IPW) tail quantile score process}. 
For $t\in[0,1]$, define
\begin{equation*}
	\mathbf{V}_{n,j}(t)
	=
	\frac{1}{\sqrt{k_n}}
	\sum_{s=1}^{n}
	\omega_{s,j}
	\left[
	\mathbb{I}
	\left\{
	Y_s
	>
	\bX_s^\top
	\bbeta_j\left(1-\frac{k_nt}{n}\right)
	\right\}
	-\frac{k_nt}{n}
	\right]
	\bX_s,
\end{equation*}
where $\omega_{s,j}=\left\{\frac{T_s}{\pi(\bX_s)}\right\}^{j}
\left\{\frac{1-T_s}{1-\pi(\bX_s)}\right\}^{1-j}$ and $j\in \{0, 1\}$.
Although the process is indexed by $t\in[0,1]$, the corresponding quantile
level $1-k_nt/n$ approaches the upper endpoint as $n\rightarrow\infty$.

The process is closely related to several stochastic processes that arise in quantile regression, causal inference, and extreme value theory. On the quantile-regression side, it is an IPW tail analogue of the ideal regression rank-score process in Equation~(5.2) of \citet{gutenbrunner1992regression}, which underlies the asymptotic theory of regression quantiles. On the causal-inference side, its weighted quantile-score structure is related to the influence-function representation of the marginal quantile treatment effect in \citet[Eq.~(6)]{Firpo:2007}; however, in our conditional quantile setting, the conditional mean of the quantile score vanishes at the true conditional quantile, so the corresponding
augmentation term is absent. The process is also related to the IPW empirical
processes used by \citet{Donald:Hsu:2014} to establish weak convergence of
potential-outcome distribution and quantile estimators. 
A key distinction from these conventional processes is that the quantile index in
$\mathbf{V}_{n,j}(t)$ moves toward the upper endpoint. In this respect, our
setting is more closely related to \citet[Theorem~3.1]{Zhang2018}, who studies
an intermediate-order quantile process indexed by a fixed multiplier of a
quantile level approaching the boundary. Our process differs, however, in that
it concerns the underlying covariate-dependent IPW quantile score rather than
the marginal quantile estimator itself.

The IPW tail quantile score process is the key stochastic object governing
the asymptotic behavior of the intermediate quantile regression estimators
$\widehat{\bbeta}_{j}(1-i_n/n)$ defined in \eqref{eq: hat_beta}.
These intermediate-level estimators, in turn, form the basis for the
extreme value index estimators $\widehat{\gamma}^{\diamond}_j$ and the
CEQTE estimators $\widehat{\Delta}^{\diamond}(p_n\mid\bx_0)$.
We therefore begin by establishing a weak convergence result for
$\mathbf{V}_{n,j}$.

\begin{theorem}\label{thm: AN_quantilescore}
	Suppose that Assumptions~\ref{asmp: Assumption 1} and~\ref{asmp: linear} hold,
	and that $k_n\rightarrow\infty$ and $k_n/n\rightarrow0$. Then, for each
	$j\in\{0,1\}$,
	\[
	\mathbf V_{n,j}
	\rightsquigarrow
	\mathcal V_j
	\qquad
	\text{in } \ell^\infty([0,1])^d,
	\]
	where $\mathcal V_j(t)$, $t\in[0,1]$, is a mean-zero $d$-dimensional Gaussian
	process with continuous sample paths and covariance function
	\[
	\operatorname{Cov}
	\left(
	\mathcal V_j(t_1),
	\mathcal V_j(t_2)
	\right)
	=
	(t_1\wedge t_2)
	E\left[
	\left\{\frac{1}{\pi(\bX)}\right\}^{j}
	\left\{\frac{1}{1-\pi(\bX)}\right\}^{1-j}
	\bX\bX^\top
	\right],
	\qquad
	t_1,t_2\in(0,1].
	\]
\end{theorem}

\begin{remark}
	
	This result holds under very general conditions and, importantly  does \textit{not} require
	$F_{Y(j)\mid\bx}$ to belong to a max-domain of attraction. Hence, it does not
	impose a particular form of tail decay and accommodates heavy-tailed,
	light-tailed, and bounded-tail conditional distributions. 
	The proof proceeds through a bracketing-number argument and the proof strategy is not inherently tied to linear quantile regression. The result can be easily extended to other parametric specifications of the conditional quantile function.
	
\end{remark}

\begin{remark}
	From an extreme-value perspective, $\mathbf{V}_{n,j}(t)$ may also be viewed
	as a covariate- and inverse-probability-weighted analogue of the classical
	uniform tail empirical process; see, for example, \cite{Einmahl1992}.
	This connection is reflected in the $\sqrt{k_n}$ normalization and the
	Brownian motion limit.
	Just like the tail empirical process, the asymptotic result of our IPW tail quantile score process can be useful in a broader context, for intance,
	for studying extreme-order linear quantile
	regression and related causal-inference procedures.
	
\end{remark}

The proof of Theorem \ref{thm: AN_quantilescore} is provided in Section S1.2 of the online Supplemental Materials. 

\subsection{Asymptotic normality  of the estimators}

We use the functional weak convergence of the IPW tail quantile score process
$\mathbf{V}_{n,j}$ to derive an asymptotic expansion for the intermediate
quantile regression estimator
$\widehat{\bbeta}_{j}(1-i_n/n)$, $i_n=m_n,\dots,k_n$. This derivation requires stronger uniform
conditions on the upper-tail quantile model. We therefore impose the following
two conditions, which strengthen Assumptions~\ref{asmp: linear} and
\ref{asmp: heavy-tail}, respectively.

\begin{assumption}
	\label{asmp: Assumption USOC}
	\leavevmode
	There exists an index $\ell_0$ such that for $\ell=0,\ldots, d-1$,
	$\lim_{\tau \rightarrow 1}\frac{\beta_{j,\ell}(\tau)}{\beta_{j,\ell_0}(\tau)}=:h_{j, \ell}\in[0,\infty)$.
	Without loss of generality, we set $\ell_0=0$ and define $\bH_j=(1,h_{j,1},\ldots, h_{j,d-1})^{\top}$. In addition, we assume that the relative magnitude of the conditional extreme quantile with respect to its marginal counterpart satisfies the following uniform approximation: \begin{eqnarray*}\label{relationship}
		\sup_{\bX\in \bD_X, m_n\leq i_n\leq k_n} \left| \frac{Q_{Y(j)|\bX}(1-\frac{i_n}{n})}{Q_{Y(j)}(1-\frac{i_n}{n})}-\bH_j^\top\bX \right| \rightarrow 0.
	\end{eqnarray*}
\end{assumption}

\begin{assumption}
	\label{asmp: Assumption SOC}
	\leavevmode
	Assume that the conditional extreme quantile 
	$Q_{Y(j)\mid\bx}(\cdot)$ satisfies the second-order regular variation 
	condition uniformly over $\bx\in D_{\bX}$, i.e.,
	\begin{align}\label{eq: usoc}
		\sup_{\bx \in \bD_X, 1/2 \leq y \leq 2}\left| \frac{
			\frac{
				Q_{Y(j)\mid\bx}\left(1-\frac{1}{vy}\right)
			}{
				Q_{Y(j)\mid\bx}\left(1-\frac{1}{v}\right)
			}
			-y^{\gamma_j}
		}{
			A_{Y(j)\mid\bx}(v)
		}
		-
		y^{\gamma_j}\frac{y^{\rho_j}-1}{\rho_j}\right| \rightarrow 0,
	\end{align}
	where $
	A_{Y(j)\mid\bx}(v)
	=
	a_j(\bx)v^{\rho_j}l_j(v),
	$
	with $a_j(\bx)>0$, $\rho_j<0$, and $l_j(\cdot)$ slowly varying.
\end{assumption}

Assumption \ref{asmp: Assumption USOC} characterizes the limiting
relationship between the conditional and marginal extreme quantiles
through $\bH_j^\top\bX$. In particular, $\bH_j^\top\bX$ takes an
explicit linear form and plays a role analogous to $K(\bz)$ in
Condition B2 of \cite{Wang2012}. 
Assumption~\ref{asmp: Assumption SOC} is a second-order strengthening of the regular
variation condition in \eqref{oneorder}. Compared with the conditions 
imposed in \citet{Wang2012}, our assumptions avoid the need to introduce an
auxiliary reference distribution and do not require differentiability of the
conditional quantile function. These conditions are therefore more
readily verifiable in applications.
Proposition~S.1 in the online Supplemental Materials further shows
that the marginal extreme quantile $Q_{Y(j)}$ inherits the same
first- and second-order indices $(\gamma_j,\rho_j)$.
All data-generating processes considered in Section \ref{Simulation}
satisfy Assumptions \ref{asmp: Assumption USOC} and
\ref{asmp: Assumption SOC}, with the corresponding parameter values
reported in Table \ref{tab:dgp_parameters}.

\begin{theorem}\label{theo: beta}
	Suppose that the estimator of the propensity score satisfies 
	$\sup\limits_{\bX\in\bD_{\bX}}\left|\frac{1}{\widehat{\pi}(\bX)}-\frac{1}{\pi(\bX)}\right|=o_p(n^{-1/4})$
	and $E\{\bX\bX^\top/(\bH_j^{\top}\bX)\}$  is a $p\times p$ positive definite matrix. 
	Assume further that $k_n\rightarrow \infty$, $k_n n^{-1/2}\rightarrow0$, $m_n/k_n\rightarrow 0$, and $m_n k_n^{-1/2}\rightarrow\infty$ as $n\rightarrow \infty$. 
	Then, under Assumptions \ref{asmp: Assumption 1}, \ref{asmp: linear}, \ref{asmp: Assumption USOC} and \ref{asmp: Assumption SOC}, we have 
	\begin{align*}
		\sup_{m_n\leq i_n\leq k_n}\left\| \frac{i_n}{\sqrt{k_n}{Q_{Y(j)}(1-\frac{i_n}{n})}}\left\{\widehat{\bbeta}_{j}\left(1-\frac{i_n}{n}\right)-\bbeta_{j}\left(1-\frac{i_n}{n}\right)\right\}-\gamma_{j}\bM_j^{-1}\mathbf{V}_{n,j}\left(\frac{i_n}{k_n}\right)  \right\| \overset{p}{\rightarrow} 0,
	\end{align*}
	as $n\rightarrow \infty$, where $\bM_j=E\{\bX\bX^\top/(\bH_j^{\top}\bX)\}$ and $\mathbf{V}_{n,j}$ is defined in Theorem \ref{thm: AN_quantilescore}.
\end{theorem}
Theorem \ref{theo: beta} establishes the uniform convergence of the intermediate quantile coefficient vector $\widehat{\bbeta}_{j}\left(1-\frac{i_n}{n}\right)$ over $m_n\leq i_n\leq k_n$, which is used to prove the asymptotic properties of the extreme value index estimator $\widehat{\gamma}^{\rm{HC}}_{j}(\bx_0)$ and the extreme quantile estimator. 
We have provided a rigorous theoretical proof of Theorem \ref{theo: beta} in Section S1.2 of the online Supplemental Materials. The propensity score estimator used in our procedure is given in Section~S2.1 of the online Supplemental Materials, and the required convergence condition is readily satisfied under mild regularity conditions; see, for example, Proposition~1 of \citet{Wang:Qin:Song:Tang:2022}.

We now establish the asymptotic properties of the proposed estimators $\widehat{\gamma}^{\rm{HC}}_{j}(\bx_0)$  and $\widehat{\Delta}^{\rm{HC}}(p_{n}|\bx_0)$. Since the estimator $\widehat{\gamma}^{\rm HM}_{j}$ is constructed similarly to that in \cite{Deuber:Li:Engelke:Maathuis:2021}, its asymptotic properties are provided in Section S1.1 of the online Supplementary Materials.

\begin{theorem}\label{theo: AN_evt}
	Suppose that the conditions of Theorem \ref{theo: beta} hold.
	If $\sqrt{k_{n}}A_{Y(j)|\bx_0}(n/k_{n})\rightarrow0$ as $n\rightarrow\infty$, then
	
	\[
	\sqrt{k_{n}}
	\begin{pmatrix}
		\widehat{\gamma}^{\rm HC}_{1}(\bx_0)-\gamma_{1} \\
		\widehat{\gamma}^{\rm HC}_{0}(\bx_0)-\gamma_{0}
	\end{pmatrix}
	\;\xrightarrow{D}\;
	\mathcal{N}\!\left(
	\begin{pmatrix}
		0 \\[0.6ex]
		0
	\end{pmatrix},
	\begin{pmatrix}
		\gamma_{1}^{2}{\sigma}^{2}_{1}(\bx_0) & 0 \\[0.6ex]
		0 & \gamma_{0}^{2}{\sigma}^{2}_{0}(\bx_0)
	\end{pmatrix}
	\right),
	\]
	where 
	\begin{align*}
		{\sigma}^{2}_{j}(\bx_0)
		=(\bH_j^{\top}\bx_0)^{-2}
		\bx_0^{\top}
		\bM_j^{-1}
		E\!\left[
		\left\{\frac{1}{\pi(\bX)}\right\}^{j}
		\left\{\frac{1}{1-\pi(\bX)}\right\}^{1-j}
		\bX\bX^{\top}
		\right]
		\bM_j^{-1}
		\bx_0,
	\end{align*}
	with $\bH_j$  defined in Assumption \ref{asmp: Assumption USOC}, $\bM_j$   defined in Theorem \ref{theo: beta}.
	
	If we further assume that $n(1-p_{n} )=o(k_{n})$, $\log\{n(1-p_{n})\}=o(\sqrt{k_{n}})$ and $\lim\limits_{p_n\rightarrow 1}\frac{Q_{Y(1)|\bx_0}(p_n)}{Q_{Y(0)|\bx_0}(p_n)}=\varpi\in[0,\infty]$. Then,
	\[
	\frac{\sqrt{k_n}}{\log\!\bigl(k_n/\{n(1-p_n)\}\bigr)}
	\cdot
	\frac{\widehat{\Delta}^{\mathrm{HC}}(p_n\mid\bx_0)-\Delta(p_n\mid\bx_0)}
	{\max\!\{Q_{Y(1)\mid\bx_0}(p_n),Q_{Y(0)\mid\bx_0}(p_n)\}}
	\xrightarrow{D}
	\mathcal{N}\!\left(0,\sigma^{2}(\bx_0)\right),
	\]
	where $\sigma^2(\bx_0)=\sigma^{2}_{0}(\bx_0)\min\{1,\varpi^{2}\}
	+
	\sigma^{2}_{1}(\bx_0)\min\{1,\varpi^{-2}\}$.
	
\end{theorem}
\begin{remark}
	A more relaxed condition on \(k_n\), allowing for a non-vanishing asymptotic bias,
	\[
	\sqrt{k_n}\,A_{Y(j)|\bx_0}\!\left(\frac{n}{k_n}\right)\longrightarrow \lambda_j \in \mathbb{R},
	\]
	can also be imposed. However, our simulation results suggest that choosing \(k_n\) of smaller order yields better finite-sample performance.
\end{remark}

\begin{remark}
	The result complements the existing asymptotic theory for extremal quantile
	treatment effects in two important respects. First, unlike
	\citet[Theorem~4.1]{Zhang2018}, which analyzes the marginal QTE at an
	extreme-order quantile without relying on EVT-based extrapolation, our
	estimator extrapolates from an intermediate quantile to the target extreme
	level using the regular variation of the potential-outcome distributions.
	Second, while \citet[Theorems 1 and 3]{Deuber:Li:Engelke:Maathuis:2021} also employ
	EVT-based extrapolation for marginal extremal QTEs, our analysis concerns
	the conditional extreme quantile treatment effect and therefore allows the
	magnitude of the treatment effect to vary with the covariates. Thus, our
	result combines EVT-based extrapolation with the analysis of heterogeneous
	treatment effects in the extreme tail.
\end{remark}

The proofs of the asymptotic results above are provided in Section S1.2 of the online Supplemental Materials.

\section{Simulation Studies}\label{Simulation}

We consider three data-generating processes (DGPs) with confounder dimensions
ranging from 2 to 6, including both discrete and continuous confounders.
In all designs, the unobserved noise component is independent of \(\bX\) and
follows either a Student’s \(t\) distribution or a Pareto distribution.
Let \(U \sim \mathrm{Uniform}(0,1)\) be independent of \(\bX\).
The DGPs are defined as follows.
\begin{align*}
	\text{Model 1:}\quad
	&\begin{cases}
		Y(0) = X_1 + X_2 + Q_{t_3}(U), \\
		Y(1) = X_1 + X_2 + (1 + 0.9 X_1)Q_{t_3}(U), \\
		\text{logit}\{\pi(\bX)\} = -1.5 X_1 - X_2,
	\end{cases}
\end{align*}
where \(X_1 \sim \mathrm{Uniform}(0,1)\) and
\(X_2 \sim \mathrm{Bernoulli}(0.8)\). The function \(Q_{t_3}\) denotes the quantile function of
\(t(3)\).
\begin{align*}
	\text{Model 2:}\quad
	&\begin{cases}
		Y(0) = \tfrac{1}{2} \sum_{\ell=1}^{5} X_\ell + (1 + 0.9 X_1)Q_{e_0}(U), \\
		Y(1) = \sum_{\ell=1}^{5} X_\ell + (1 + 0.9 X_1)Q_{e_1}(U), \\
		\text{logit}\{\pi(\bX)\} = -1.5 X_1 - \sum_{\ell=2}^{5} X_\ell,
	\end{cases}
\end{align*}
where \(X_\ell \sim \mathrm{Uniform}(0,1)\) for \(\ell=1,\dots,4\), \(X_5 \sim \mathrm{Bernoulli}(0.8)\).
The functions \(Q_{e_0}\) and \(Q_{e_1}\) denote the quantile functions of
Pareto\((0.2)\) and Pareto\((0.5)\), respectively.
\begin{align*}
	\text{Model 3:}\quad
	&\begin{cases}
		Y(0) = X_1 + Q_{t_3}(U), \\
		Y(1) = b(U) X_1 + Q_{t_3}(U), \\
		\text{logit}\{\pi(\bX)\} = 0.75 - 1.5 X_1,
	\end{cases}
\end{align*}
where \(X_1 \sim \mathrm{Beta}(2,1)\) 
and $b(U)=1+0.9\bigl\{Q_{t_3}(U)-Q_{t_3}(0.9)\bigr\}\mathbb{I}(0.9\le U <1)
+0.9\bigl\{Q_{t_3}(0.1)-Q_{t_3}(U)\bigr\}\mathbb{I}(0<U<0.1).$

These DGPs are designed to represent distinct forms of heterogeneity in
conditional quantile treatment effects.
Model~1 induces confounder-dependent heteroskedasticity.
Model~2 allows for distributional heterogeneity through
different tail behaviors of \(Y(0)\) and \(Y(1)\).
Model~3 generates localized treatment effects that arise
exclusively in the lower and upper tails of the outcome distribution. All three models satisfy Assumptions \ref{asmp: Assumption USOC} and \ref{asmp: Assumption SOC}. The parameters  $(\gamma_j, \rho_j)$ and $\bH_j$, and the implied conditional treatment-effect quantities are summarized in Table \ref{tab:dgp_parameters}.

\begin{landscape}
	\begin{figure}[htbp]
		
		\centering
		\includegraphics[width=1.1\textwidth,height=1.0\textheight]{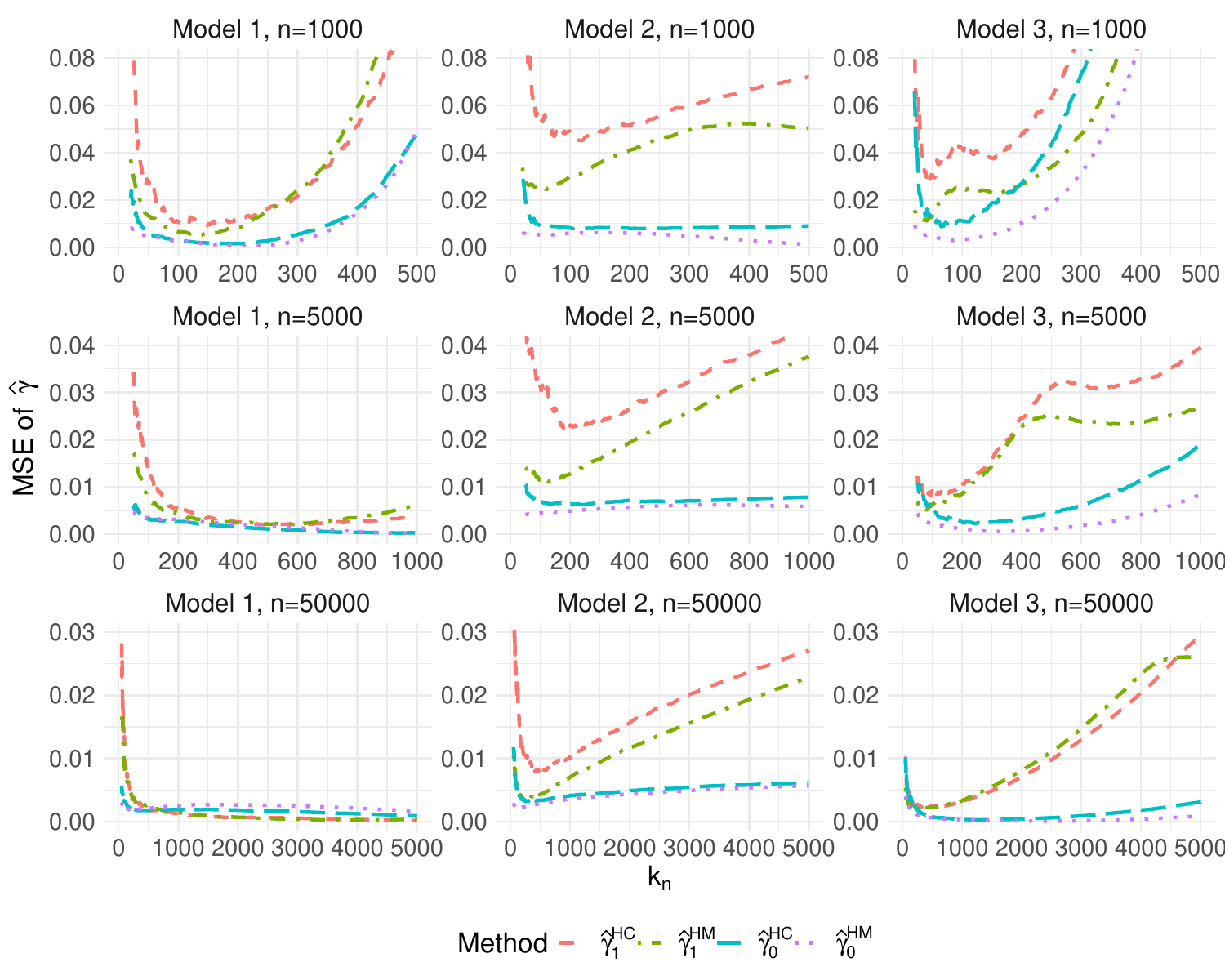}
		\caption{MSE of $\widehat{\gamma}^\diamond_j$ for varying $k_n$ across three models and three sample sizes $n=1000, 5000, 50000$.}\label{Fig: MSE of gamma_1}
	\end{figure}
\end{landscape}

For all three models, we consider three sample sizes \(n=1{,}000\), \({5,}000\) and \(50{,}000\),
with 100 Monte Carlo replications in each case. We first examine the performance
of the two proposed estimators of the extreme value index (EVI): the
confounder-dependent estimator \(\widehat{\gamma}^{\mathrm{HC}}_{j}(\bx_0)\)
defined in~\eqref{eq: hatgamma1} and the marginal estimator
\(\widehat{\gamma}^{\mathrm{HM}}_{j}\) defined in~\eqref{eq: hatgamma2}. 
For the confounder-dependent estimator, we set
\(\bx_0^\top=(0.3,1)\), \((0.8,0.8,0.8,0.8,1)\), and \((0.5)\) for Models~1--3,
respectively, and fix \(m_n=10\) in all cases. 
Figure~\ref{Fig: MSE of gamma_1}  display the mean squared errors (MSE) of
the two estimators as $k_n$ changes. We make three observations based on these results.
\begin{itemize}
\item The HM estimator generally yields smaller MSEs over a broad range of $k_n$, illustrating the benefit of pooling tail information across covariate values through the marginal distribution rather than relying on increasingly extreme conditional quantile estimates. This advantage becomes particularly pronounced as the covariate dimension increases. In Model~(2), where $d=5$, HM continues to outperform HC even for very large sample sizes. These results suggest that the marginal estimator is particularly attractive in multivariate covariate settings, where stable estimation of extreme conditional quantiles can be difficult.

\item In low-dimensional settings ($d=1$ or $2$), the HC estimator becomes increasingly competitive and, for sufficiently large samples, eventually outperforms HM. This pattern highlights a dimension--sample-size trade-off between the two estimators: pooling across covariate values improves stability when tail information is limited, whereas direct estimation from conditional quantiles can be more efficient when the covariate dimension is low and sufficient data are available.

\item At the selected values of $k_n$, the MSEs for the control group are generally smaller than those for the treated group. Allowing group-specific intermediate levels may therefore further improve finite-sample performance when the potential-outcome distributions exhibit different tail behaviors or when the treatment groups are substantially imbalanced.
\end{itemize}

Next, we examine the estimation of the conditional extreme quantile treatment effect
\(\Delta(p_n\mid\bx_0)\), where \(p_n=1-rn^{-1}\) with \(r\in\{10,5,0.5\}\).
In addition to the two proposed estimators in~\eqref{eq: hatDelta}, we consider
two competing approaches. The first competitor is the estimator of \citet{Deuber:Li:Engelke:Maathuis:2021},
given by
\[
\widehat{\Delta}^{\mathrm{MM}}(p_n\mid\bx_0)
=
\Big\{\tfrac{k_n}{n(1-p_n)}\Big\}^{\widehat{\gamma}^{\mathrm{HM}}_{1}}
\widetilde{Q}_{Y(1)}\!\left(1-\tfrac{k_n}{n}\right)
-
\Big\{\tfrac{k_n}{n(1-p_n)}\Big\}^{\widehat{\gamma}^{\mathrm{HM}}_{0}}
\widetilde{Q}_{Y(0)}\!\left(1-\tfrac{k_n}{n}\right),
\]
where \(\widehat{\gamma}^{\mathrm{HM}}_{j}\) is defined in~\eqref{eq: hatgamma2}
and \(\widetilde{Q}_{Y(j)}(1-k_n/n)\) is defined in~\eqref{eq: QYj-uncond}.
This estimator targets the marginal extreme quantile treatment effect and
therefore does not depend on \(\bx_0\). 

The second competitor is the inverse-probability-weighted quantile regression
estimator of \citet{Firpo:2007},
\[
\widehat{\Delta}^{\mathrm{QR}}(p_n\mid\bx)
=
\bx^{\top}\{\widehat{\bm\beta}_{1}(p_n)-\widehat{\bm\beta}_{0}(p_n)\},
\]
where $\widehat{\bm\beta}_{j}$ is given by \eqref{eq: hat_beta}.

For all methods, the propensity score \(\widehat{\pi}(\bX_i)\) is estimated using
a multiply robust procedure; see Section~S2.1 of the online Supplementary
Materials.

\begin{landscape}
	\begin{figure}[htbp]
		\centering
		\includegraphics[width=1.5\textwidth,height=0.8\textheight]{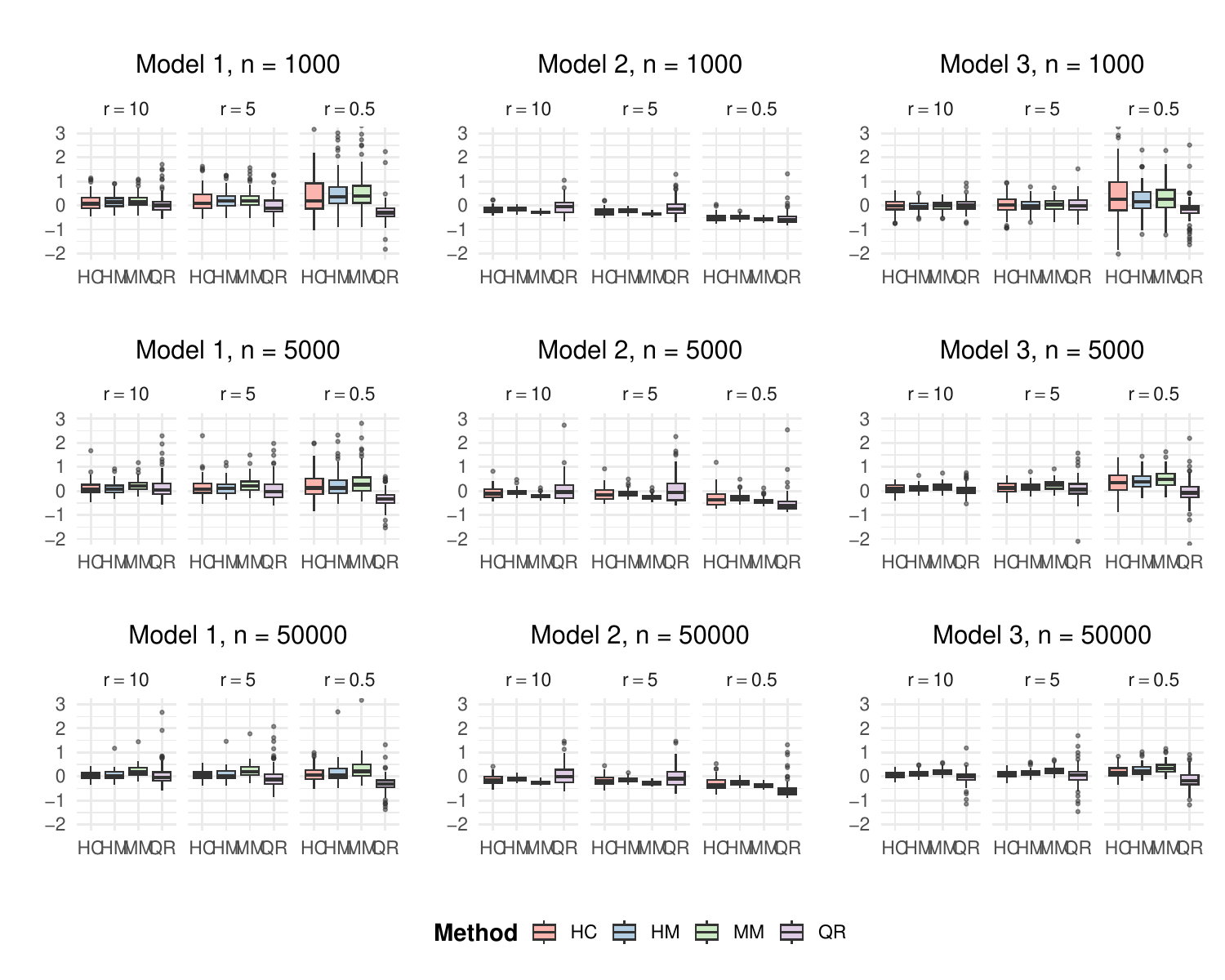}
		\caption{Boxplot of the normalized estimation error of $\widehat{\Delta}^{\diamond}(p_n\mid\bx_0)$, where  \(p_n=1-rn^{-1}\) with \(r\in\{10,5,0.5\}\), and \(\bx_0^\top=(0.3,1)\), \((0.8{\bf 1}_4^\top,1)\), and \((0.5)\) respectively for Models~1--3. The same $k_n$ is used for EVI 
and intermediate quantile estimation, with $k_n=(200,350,200)$ 
for Models~1--3 when $n\in\{1{,}000,5{,}000\}$ and 
$k_n=(500,1{,}000,500)$ when $n=50{,}000$.}
		\label{Fig: Boxplot of scaled error}
	\end{figure}
\end{landscape}

To compare the four methods, we consider the normalized estimation error
\[
\widetilde{E}^{\,\diamond}(p_n,\bx_0)
=
\frac{\widehat{\Delta}^{\diamond}(p_n\mid\bx_0)-\Delta(p_n\mid\bx_0)}
{\max\{Q_{Y(1)\mid\bx_0}(p_n),\,Q_{Y(0)\mid\bx_0}(p_n)\}},
\qquad
\diamond\in\{\mathrm{HC},\mathrm{HM},\mathrm{MM},\mathrm{QR}\}.
\]
Figure~\ref{Fig: Boxplot of scaled error} displays boxplots of the normalized
estimation error based on 100 Monte Carlo replications. For Figure~\ref{Fig: Boxplot of scaled error}, we use the same $k_n$ for EVI and intermediate quantile 
estimation. Specifically, we set $k_n=200$ for Models~1 and~3 and $k_n=350$ for Model~2 when $n\in\{1{,}000,5{,}000\}$; when $n=50{,}000$, the corresponding values are $500$ and $1{,}000$, respectively. For the smaller sample size, $n=1{,}000$, estimation is challenging for all four methods, with relatively large bias and variability, particularly when $r=0.5$. The QR estimator systematically underestimates the CEQTE in this case, as $r=0.5$ corresponds to a target quantile beyond the effective sample range and therefore requires extrapolation. Among the three EVT-based methods, MM and HM exhibit similar variability and are generally more stable than HC for $n=1{,}000$ and $5{,}000$. However, MM consistently exhibits larger bias than HM, suggesting that directly targeting the conditional extreme quantile treatment effect better captures treatment-effect heterogeneity than relying on marginal extreme quantile treatment effects.  

To facilitate an overall comparison across methods, we use the relative mean absolute error (rMAE), taking the QR estimator as the benchmark:
\begin{eqnarray}
	\text{rMAE}^{\,\diamond}(p_n,\bx_0)=\frac{\sum_{b=1}^{B}\left|{\widehat{\Delta}^{\diamond}_{b}(p_n|\bx_0)-\Delta(p_n|\bx_0)}\right|}
	{\sum_{b=1}^{B}\left|{\widehat{\Delta}^{\mathrm{QR}}_{b}(p_n|\bx_0)-\Delta(p_n|\bx_0)}\right|}, \qquad
	\diamond\in\{\mathrm{HC},\mathrm{HM},\mathrm{MM}\},
	\label{eq: rMAE}
\end{eqnarray}
where $B=100$ denotes the number of Monte Carlo replications. Thus, an rMAE below one indicates that the corresponding method outperforms the QR benchmark in terms of mean absolute error. 
The rMAE results are reported in Table~\ref{Tab: MAE_rel}, where we use the same $k_n$ as in Figure~\ref{Fig: Boxplot of scaled error} for EVI 
estimation and consider two values of $k_n$ for intermediate quantile 
estimation at each sample size: $(100,200)$ for $n=1{,}000$, 
$(200,350)$ for $n=5{,}000$, and $(500,1{,}000)$ for $n=50{,}000$.
\begin{itemize}
\item For the small and intermediate sample sizes, $n=1{,}000$ and $5{,}000$, HM generally performs best among the four methods. Its advantage becomes more pronounced as the covariate dimension increases. In particular, HM attains the smallest rMAE in Model~2 ($d=5$), which is consistent with the greater stability of the marginal EVI estimator when conditional tail information is sparse across multiple covariate dimensions.
\item QR outperforms the EVT-based estimators in several settings in Models~1 and~3, especially when the sample size is limited and the target quantile is particularly extreme. The EVT-based methods rely on both estimated intermediate conditional quantiles and estimated EVIs for extrapolation, so estimation errors in either component can be magnified when estimating the CEQTE at very extreme levels. This highlights the importance of a careful selection of the tuning parameter $k_n$.
As for MM, it generally underperforms the other methods, with only isolated exceptions in Model~3.
\item When the sample size increases to $n=50{,}000$, HC becomes more competitive and outperforms HM in the low-dimensional settings ($d=1$ or $2$).
\end{itemize}
As a practical recommendation, HM provides a robust default choice across a broad range of settings, whereas HC may be preferred when the covariate dimension is low and the sample size is sufficiently large.

\begin{landscape}
	\begin{table}[H]
		\footnotesize
		\centering
		\caption{Model parameters and implied conditional treatment-effect quantities.}
		\label{tab:dgp_parameters}
		
		\setlength{\tabcolsep}{6pt}
		\renewcommand{\arraystretch}{1.25}
		
		\begin{tabular}{l l l c c l}
			\toprule
			\textbf{DGP}
			& \makecell[l]{$\bbeta_0^\top(\tau)$\\ $\bbeta_1^\top(\tau)$}
			& \makecell[l]{$\bH_0^\top$\\ $\bH_1^\top$}
			& \makecell[l]{$(\gamma_0,\rho_0)$\\ $(\gamma_1,\rho_1)$}
			& \textbf{CATE}
			& \makecell[l]{$\tau$-CQTE}
			\\
			\midrule
			
			\textbf{Model 1}
			& \makecell[l]{$(Q_{t_3}(\tau),1,1)$\\ $(Q_{t_3}(\tau),(1+0.9)Q_{t_3}(\tau),1)$}
			& \makecell[l]{$(1,0,0)$\\ $(1,0.9,0)$}
			& \makecell[l]{$(\tfrac13,-\tfrac13)$\\ $(\tfrac13,-\tfrac13)$}
			& $0$
			& $0.9\,Q_{t_3}(\tau)\,x_1$
			\\
			\addlinespace[0.6ex]
			\midrule
			\addlinespace[0.2ex]
			
			\textbf{Model 2}
			& \makecell[l]{$(Q_{e_0}(\tau),\tfrac12+0.9Q_{e_0}(\tau),\tfrac12\bf{1}_4)$\\
				$(Q_{e_1}(\tau),1+0.9Q_{e_1}(\tau),\bf{1}_4)$}
			& \makecell[l]{$(1,0.9,\bf{0}_4)$\\ $(1,0.9,\bf{0}_4)$}
			& \makecell[l]{$(\tfrac12,-\tfrac12)$\\ $(\tfrac15,-\tfrac15)$}
			& $0.75(1+0.9x_1)+0.5\sum\limits_{\ell=1}^{5}x_\ell$
			& $\{Q_{e_1}(\tau)-Q_{e_0}(\tau)\}(1+0.9x_1)+0.5\sum\limits_{\ell=1}^{5}x_\ell$
			\\
			\addlinespace[0.6ex]
			\midrule
			\addlinespace[0.2ex]
			
			\textbf{Model 3}
			& \makecell[l]{$(Q_{t_3}(\tau),1)$\\ $(Q_{t_3}(\tau), b(\tau))$}
			& \makecell[l]{$(1,0)$\\ $(1,0.9)$}
			& \makecell[l]{$(\tfrac13,-\tfrac13)$\\ $(\tfrac13,-\tfrac13)$}
			& $0$
			& $\left\{
			\begin{aligned}
				&\{-0.9Q_{t_3}(0.9)+0.9Q_{t_3}(\tau)\}x_1, &&0.9\le\tau<1,\\
				&0, &&0.1\le\tau<0.9,\\
				&\{-0.9Q_{t_3}(\tau)+0.9Q_{t_3}(0.1)\}x_1, &&0<\tau<0.1
			\end{aligned}
			\right.$
			\\
			
			\bottomrule
		\end{tabular}
		
		\vspace{1.0mm}
		\begin{flushleft}
			\footnotesize
			$\bf{1}_4$ and $\bf{0}_4$ denote 4-dimensional row vectors of ones and zeros, respectively.
		\end{flushleft}
		
	\end{table}
\end{landscape}


\begin{table}[htbp]
	\centering
	\caption{Relative  mean absolute  error of $\widehat{\Delta}^{\diamond}(p_n\mid\bx_0)$ defined in \eqref{eq: rMAE}, where  \(p_n=1-rn^{-1}\) with \(r\in\{10,5,0.5\}\), and \(\bx_0^\top=(0.3,1)\), \((0.8\times {\bf 1}_4^\top,1)\), and \((0.5)\) respectively for Models~1--3. }
	\label{Tab: MAE_rel}
	\renewcommand{\arraystretch}{1.0}
	\begin{tabular}{lllccc|ccc|ccc}
		\hline
		&&& \multicolumn{3}{c|}{Model 1} & \multicolumn{3}{c|}{Model 2} & \multicolumn{3}{c}{Model 3} \\
		\cmidrule(lr){4-6}\cmidrule(lr){7-9}\cmidrule(lr){10-12}
		$n$&$k_n$& $r$ & HC & HM & MM & HC & HM & MM & HC & HM & MM\\
		\hline
		\multirow{6}{*}{1000}
		& \multirow{3}{*}{100} 
		& 10  &1.00 	&\textbf{0.82} 	&1.09 	&0.79 	&\textbf{0.67} 	&1.16 	&1.34 	&0.93 	&\textbf{0.92} \\
		&     & 5   &0.93 	&\textbf{0.75} 	&0.98 	&0.88 	&\textbf{0.77} 	&1.16 	&1.31 	&\textbf{0.85} 	&0.87 \\
		&     & 0.5 &1.99 	&1.59 	&1.95 	&0.90 	&\textbf{0.85} 	&1.00 	&2.41 	&1.33 	&1.48 \\
		\cmidrule(lr){3-12}
		& \multirow{3}{*}{200} 
		& 10  &0.93 	&\textbf{0.80} 	&0.89 	&0.81 	&\textbf{0.66} 	&1.17 	&1.20 	&\textbf{0.83} 	&0.86  \\
		&     & 5   &0.87 	&\textbf{0.74} 	&0.82 	&0.90 	&\textbf{0.77} 	&1.17 	&1.21 	&\textbf{0.78} 	&0.85 \\
		&     & 0.5 &1.88 	&1.57 	&1.73 	&0.91 	&\textbf{0.85} 	&1.01 	&2.33 	&1.29 	&1.50  \\
		\hline
		\multirow{6}{*}{5000}
		& \multirow{3}{*}{200}
		& 10  &0.66 	&\textbf{0.59} 	&0.80 	&0.63 	&\textbf{0.39} 	&0.72 	&1.05 	&\textbf{0.91} 	&1.13  \\
		&     & 5   &0.80 	&\textbf{0.70} 	&0.91 	&0.58 	&\textbf{0.39} 	&0.67 	&0.80 	&\textbf{0.71} 	&0.86  \\
		&     & 0.5 &1.16 	&\textbf{0.94} 	&1.15 	&0.64 	&\textbf{0.52} 	&0.71 	&1.44 	&1.30 	&1.52  \\
		\cmidrule(lr){3-12}
		& \multirow{3}{*}{350}
		& 10   &0.75 	&\textbf{0.63} 	&0.82 	&0.62 	&\textbf{0.37} 	&0.68 	&1.12 	&\textbf{0.90} 	&1.04  \\
		&     & 5  &0.89 	&\textbf{0.73} 	&0.93 	&0.57 	&\textbf{0.37} 	&0.64 	&0.82 	&\textbf{0.69} 	&0.79 \\
		&     & 0.5 &1.25 	&\textbf{0.97} 	&1.19 	&0.63 	&\textbf{0.50} 	&0.70 	&1.43 	&1.25 	&1.39  \\
		\hline
		\multirow{6}{*}{50000}
		& \multirow{3}{*}{500}
		& 10  &\textbf{0.49} 	&0.53 	&0.84 	&0.63 	&\textbf{0.31} 	&0.53 	&\textbf{0.62} 	&0.66 	&1.05  \\
		&     & 5   &\textbf{0.45} 	&0.48 	&0.73 	&0.65 	&\textbf{0.32} 	&0.55 	&\textbf{0.51} 	&0.55 	&0.83  \\
		&     & 0.5 &\textbf{0.59} 	&0.64 	&0.85 	&0.54 	&\textbf{0.30} 	&0.45 	&\textbf{0.83} 	&0.91 	&1.22 \\
		\cmidrule(lr){3-12}
		& \multirow{3}{*}{1000}
		&10	&\textbf{0.58} 	&0.61 	&0.91 	&0.54 	&\textbf{0.33} 	&0.73 	&0.68 	&\textbf{0.67} 	&0.99 \\
		& &5	&\textbf{0.52} 	&0.55 	&0.79 	&0.57 	&\textbf{0.39} 	&0.75 	&\textbf{0.54} 	&0.55 	&0.79 \\
		&&0.5	&\textbf{0.66} 	&0.70 	&0.92 	&0.50 	&\textbf{0.42} 	&0.61 	&\textbf{0.84} 	&0.90 	&1.18 \\
		\hline
	\end{tabular}
	\vspace{1.0mm}
	\begin{flushleft}
		\footnotesize The reported $k_n$ values are for intermediate quantile estimation.
For EVI estimation, $k_n=(200,350,200)$ for Models~1--3 when
$n\in\{1{,}000,5{,}000\}$ and $(500,1{,}000,500)$ when $n=50{,}000$.
Values below one indicate improvement over QR; the smallest value in each
column is in bold.
	\end{flushleft}
\end{table}

\clearpage

\section{The CEQTE of College Education on Hourly Wages}\label{Real data}

We apply the proposed framework to estimate the causal effect of college education at the tail of hourly wages. A substantial literature has examined the causal effect, see for examples, \cite{card1999causal, heckman2015causal, heckman2018returns, zhou2024attendance}. 
Among these, \citet{Deuber:Li:Engelke:Maathuis:2021} studied extreme quantile treatment effect of college education on wages. Their analysis, however, focuses on marginal treatment effects, with confounders incorporated only through the estimation of the propensity score. In contrast, the estimated CEQTE provides the quantification of causal effects across different subpopulations.

We use the same data set as \citet{Heckman2006} and \citet{Deuber:Li:Engelke:Maathuis:2021}, drawn from the National Longitudinal Survey of Youth 1979 (NLSY79), which comprises a representative sample of 805 young Americans aged 14--21 at their first interview in 1979 and is available at \url{https://www.journals.uchicago.edu/doi/suppl/10.1086/698760}. The survey contains rich information on education, labor-market outcomes, parental background, cognitive test scores, and behavioral measures. The outcome variable \(Y\) is the hourly wage (in U.S.\ dollars) at age 30, and the treatment indicator \(T\) denotes whether an individual has received a college education.
The initial set of confounders includes 19 variables capturing race, region of residence in 1979 (Hispanic, South, West, Northeast), urban status in 1979, broken home statue, age in 1979, number of siblings, family income in 1979, parental education, Armed Services Vocational Aptitude Battery (ASVAB) test scores (six components), and grade point averages in ninth-grade core subjects; see \citet{Deuber:Li:Engelke:Maathuis:2021} and references therein for a complete description. 

To present a more clear picture of the heterogeneous treatment effect, we first apply a LASSO-type approach to select possibly important confounders, while the treatment variable \(T\) is retained without penalization. 
Specifically, we implement LASSO followed by adaptive LASSO at an intermediate quantile level 0.95 (close to $1-k_n/n$ with $k_n$ selected in EVI estimation), where the tuning parameters are selected via BIC. This procedure keeps \(T\) and three confounders: \(X_1\) (binary, 1 for Northeast residence in 1979, 0 otherwise), \(X_2\) (continuous, family income in 1979), and \(X_3\) (continuous, ASVAB component~1). Then we rescale \(X_2\) and \(X_3\) to [0,1] by min--max normalization, and apply the proposed framework to estimate the CEQTE across different confounder profiles. In particular, we consider $2\times3\times3=18$ profiles, $(X_1,Q_{X_2}(p_2),Q_{X_3}(p_3))^\top$, where $X_1=0$ or 1, $p_2,p_3\in \{0.3, 0.5, 0.7\}$.

\begin{figure}[htbp]
	\centering	\includegraphics[width=1.0\textwidth]{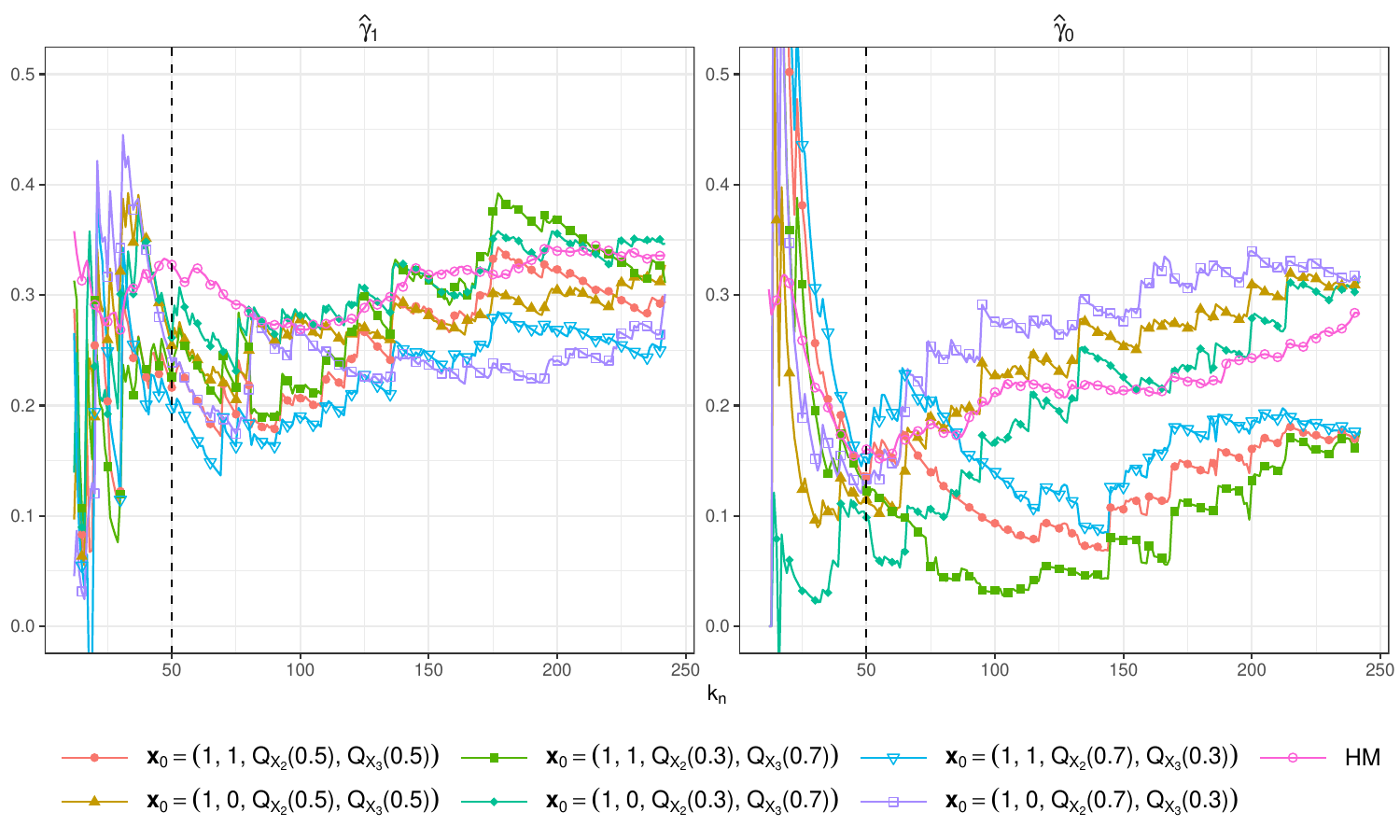}
	\caption{Estimates of extreme value indices as functions of $k_n$. Left: $\widehat{\gamma}^\diamond_1$; Right: $\widehat{\gamma}^\diamond_0$.
		\protect\ggcircle: HM estimator; other shapes: HC estimators conditional on different $\bx_0$. }\label{RD-gamma}
\end{figure}

Next we estimate the EVI. Figure~\ref{RD-gamma} displays the two proposed Hill estimators, HC defined in equation~\eqref{eq: hatgamma1} and HM defined in equation~\eqref{eq: hatgamma2}, as functions of \(k_n\). For the HC estimator, the curves are reported for several confounder values. Following the common heuristic of selecting the first region of stability, we set \(k_n = 50\) for both the treatment and control groups, and the corresponding estimated EVIs are \(0.25\) and \(0.15\), respectively.

Based on the intermediate quantile level $1-k_n/n\approx0.938$, we apply the proposed method to estimate the CEQTE \(\widehat{\Delta}^{\diamond}(p_n \mid \bx_0)\) defined in \eqref{eq: hatDelta}, at extreme quantile levels $p_n\in\{0.99,0.992,0.995,0.999\}$, across 18 confounder profiles, which are summarized in  Table~\ref{real-CEQTE}.
These estimates characterize the effect of college education on extremely high hourly wages across different subpopulations, reflecting three main observations. First, the CEQTE increases as $p_n$ increases for each subpopulation (fixed $\bx_0$), this is due to the fact that the hourly income of the treatment group has a heavier tail than that of the control group, that is, $\widehat \gamma_0<\widehat \gamma_1$. This observation is also noted in \cite{Deuber:Li:Engelke:Maathuis:2021} for the marginal extreme quantile treatment effect.
Second, the effect of college education is heterogeneous across subpopulations. 
Holding \((x_2, x_3)\) fixed, the estimated effect is substantially larger for individuals residing in the Northeast (\(x_1=1\)) than for those who do not (\(x_1=0\)), with the difference consistently exceeding 7 dollars per hour across all combinations of \((x_2, x_3)\). The effect also increases by more than 12 dollars per hour—over 50\% in relative terms—as the ASVAB component~1 score rises from the 30th to the 70th percentile. In contrast, the effect varies only minimally with family income when the other confounders are held fixed. Third, for each \(p_n\), the CEQTE differs markedly from the marginal extreme quantile treatment effect \citep[EQTE, the estimator of][]{Deuber:Li:Engelke:Maathuis:2021} reported in the last column of the table. While the CEQTE varies across confounder quantiles and reveals pronounced heterogeneity in the effect of college education, the marginal estimates provide only a benchmark and thus mask such heterogeneity.

\begin{table}[!ht]
	\centering
	\caption{Estimates of the conditional extreme quantile treatment effects (CEQTE) at $p_n=0.99,0.992,0.995,0.999$ with $k_n=50$. }\label{real-CEQTE}
	\renewcommand{\arraystretch}{1.0}
	\resizebox{1.0\textwidth}{!}{
		\begin{tabular}{c c c ccc ccc|c}
			\toprule
			&  &  & \multicolumn{3}{c}{$x_1=1$} & \multicolumn{3}{c}{$x_1=0$} \\
			\cmidrule(lr){4-6} \cmidrule(lr){7-9}
			$p_n$ & \diagbox{$x_2$}{ $x_3$} &  & $Q_{X_3}(0.3)$ & $Q_{X_3}(0.5)$ & $Q_{X_3}(0.7)$ 
			& $Q_{X_3}(0.3)$ & $Q_{X_3}(0.5)$ & $Q_{X_3}(0.7)$&EQTE \\
			\midrule
			\multirow{3}{*}{0.99}& $Q_{X_2}(0.3)$ & &17.28 	&23.81 	&30.34 	&9.59 	&16.12 	&22.66  &\multirow{3}{*}{18.13}\\
			& $Q_{X_2}(0.5)$ & &17.26 	&23.79 	&30.32 	&9.57 	&16.10 	&22.64 \\
			& $Q_{X_2}(0.7)$ & &17.25 	&23.78 	&30.31 	&9.56 	&16.09 	&22.62  \\
			\cmidrule(lr){4-10}
			\multirow{3}{*}{0.992} & $Q_{X_2}(0.3)$ & &18.92 	&25.84 	&32.76 	&10.83 	&17.75 	&24.67 &\multirow{3}{*}{19.93} \\
			& $Q_{X_2}(0.5)$ & &18.95 	&25.87 	&32.79 	&10.86 	&17.79 	&24.71 \\
			& $Q_{X_2}(0.7)$ & &18.97 	&25.89 	&32.81 	&10.88 	&17.80 	&24.73  \\
			\cmidrule(lr){4-10}
			\multirow{3}{*}{0.995} & $Q_{X_2}(0.3)$ & &22.75 	&30.57 	&38.39 	&13.75 	&21.57 	&29.39 &\multirow{3}{*}{24.16} \\
			& $Q_{X_2}(0.5)$ & &22.92 	&30.74 	&38.56 	&13.92 	&21.74 	&29.56  \\
			& $Q_{X_2}(0.7)$ & &23.01 	&30.83 	&38.65 	&14.01 	&21.83 	&29.65  \\
			\cmidrule(lr){4-10}
			\multirow{3}{*}{0.999} & $Q_{X_2}(0.3)$ & &40.86 	&52.71 	&64.56 	&27.82 	&39.67 	&51.53 &\multirow{3}{*}{44.21} \\
			& $Q_{X_2}(0.5)$ & &41.70 	&53.55 	&65.41 	&28.66 	&40.52 	&52.37  \\
			& $Q_{X_2}(0.7)$ & &42.14 	&54.00 	&65.85 	&29.11 	&40.96 	&52.81  \\
			\bottomrule
	\end{tabular}}
\end{table}

\section{Discussion}\label{Discussion}

This paper develops a framework for causal inference on heterogeneous treatment effects at extreme quantiles under heavy-tailed outcomes. A central structural feature of the linear conditional quantile model is that the conditional and marginal distributions of each potential outcome share a common extreme value index (EVI). This property gives rise to two complementary approaches to tail-index estimation and, importantly, enables marginal information to stabilize estimation precisely when conditional tail data are scarce. The main theoretical contribution is the inverse probability weighted (IPW) tail quantile score process, whose weak convergence underpins the asymptotic validity of the intermediate quantile, EVI, and ultimately CEQTE estimators. More broadly, this process offers a versatile building block for studying weighted tail quantile regression and other tail-sensitive causal problems involving treatment assignment or missing outcomes, extending the reach of our methodology beyond the present setting.

Several promising avenues remain for future work. First, the common-EVI property, which is central to our identification strategy, warrants investigation beyond linear conditional quantile models. Understanding how this property manifests—or can be relaxed—under nonlinear or semiparametric specifications would open the door to more flexible CEQTE estimation while preserving the efficiency gains from marginal tail information. Second, the presence of high-dimensional or multimodal covariates introduces both practical and theoretical challenges; recovering tail heterogeneity in such settings may require regularization, dimension reduction, or structured latent representations. Third, the interplay between limited treatment overlap and tail sparsity is particularly acute, since extreme propensity scores can drastically reduce the effective sample size in already data-sparse tail regions. 
Finally, a practically important direction is the data-adaptive selection of intermediate tail thresholds, along with inference procedures that properly account for this selection—a step that is essential for reliable finite-sample performance across diverse applications. Addressing these interconnected challenges would significantly broaden the applicability and robustness of extreme quantile treatment effect analysis.

\section{Disclosure statement}\label{disclosure-statement}

The authors declare that no conflicts of interest exist.

\section{Supplementary Materials}

The online Supplementary Materials include all the technical proofs and some implementation details.

\section*{Acknowledgments}

Generative AI tool ChatGPT (GPT-4) was occasionally used for language editing and proofreading.



%
%
%
%
%

  \bibliography{bibliography.bib}

\end{document}